# Entropy estimation of symbol sequences

Thomas Schürmann and Peter Grassberger
*Department of Theoretical Physics, University of Wuppertal, D-42097 Wuppertal, Germany*

We discuss algorithms for estimating the Shannon entropy $h$ of finite symbol sequences with long range correlations. In particular, we consider algorithms which estimate $h$ from the code lengths produced by some compression algorithm. Our interest is in describing their convergence with sequence length, assuming no limits for the space and time complexities of the compression algorithms. A scaling law is proposed for extrapolation from finite sample lengths. This is applied to sequences of dynamical systems in non-trivial chaotic regimes, a 1-D cellular automaton, and to written English texts.

**Partially random chains of symbols $s_1, s_2, s_3, \ldots$ drawn from some finite alphabet (we restrict ourselves here to finite alphabets though most of our considerations would also apply to countable ones) appear in practically all sciences. Examples include spins in one-dimensional magnets, written texts, DNA sequences, geological records of the orientation of the magnetic field of the earth, and bits in the storage and transmission of digital data. An interesting question in all these contexts is to what degree these sequences can be ''compressed'' without losing any information. This question was first posed by Shannon[1] in a probabilistic context. He showed that the relevant quantity is the *entropy* (or average information content) $h$, which in the case of magnets coincides with the thermodynamic entropy of the spin degrees of freedom. Estimating the entropy is non-trivial in the presence of complex and long range correlations. In that case one has essentially to understand perfectly these correlations for optimal compression and entropy estimation, and thus estimates of $h$ measure also the degree to which the structure of the sequence is understood.**

## I. INTRODUCTION

Partially random chains of symbols $s_1, s_2, s_3, \ldots$ drawn from some finite alphabet appear in practically all sciences (we restrict ourselves here to finite alphabets though most of our considerations would also apply to countable ones). We might just mention spins in 1-dimensional magnets, written texts, DNA sequences, geological records of the orientation of the magnetic field of the earth, and bits in the storage and transmission of digital data. An interesting question in all these contexts is to what degree these sequences can be ''compressed'' without losing any information. This question was first asked by Shannon[1] in a probabilistic context. He showed that the relevant quantity is the *entropy* (or average information content) $h$ which in the case of magnets coincides with the thermodynamic entropy of the spin degrees of freedom.

Another application which we are particularly interested in is chaotic dynamical systems. Assume we have a time series $x_t$, $t=1,\ldots,N$, where time is discretized but $x_t$ is continuous. In order to reduce this to the above case, we discretize $x_t$ by defining some partition $\mathcal{P}_\epsilon$ in the phase space where all elements have diameter $<\epsilon$. We represent the time series by the string $s_1, s_2 \ldots, s_t, \ldots$, where $s_t = \sigma$ means that $x_t$ is in the $\sigma$-th element of $\mathcal{P}_\epsilon$. This induces what is called a ''symbolic dynamics.'' It was pointed out by Kolmogorov and Sinai[2,3] that the entropy of such a symbol string converges for $\epsilon \to 0$ to a finite non-zero value $h_{KS}$ (the KS or ''metric'' entropy) iff the system generating the time series is chaotic. Thus measuring the entropy of a symbolic dynamics is important for deciding whether the system is chaotic.

Moreover, it was shown in Refs. 2, 3 that it is in general not necessary to take the limit $\epsilon \to 0$: there exist ''generating'' partitions (eventually infinite but countable) whose entropy is exactly $h_{KS}$. But for most chaotic systems such generating partitions are not explicitly known. The best known exception are $1-d$ maps where any partition into monotonic laps is generating. Already for the Hénon map $(x,y) \to (1.4 - x^2 + 0.3y, x)$ no rigorous construction of a generating partition exists, though there are strong indications[4] that a heuristic argument based on homoclinic tangencies[5] gives a correct result.

Even if we do not have any problems of finding a good partition, estimating the entropy can be highly non-trivial. This is always true if there are strong long range correlations. Such correlations can help to achieve higher compression rates (since they reduce the uncertainty of yet unseen symbols). But finding them and taking them into account can be very difficult because of the exponential increase of the number of different blocks (or ''words'') of symbols with the block length.

For natural languages this can be partially overcome by means of subjective methods[6] which use the fact that humans know the structure of their own language sufficiently well to be able to guess most of the information it provides on single missing letters. Thus they often can guess letters even when provided with less information than a machine would need. Algorithms based on this idea have been improved considerably,[7,8] but they are still unreliable, slow, and restricted to natural languages.

The most straightforward objective method consists in counting the frequencies of all blocks up to a certain length and estimating from them their probabilities. For an alphabet with $d$ symbols this usually breaks down when $d^n \approx N$ (where $n =$ block length) which gives, e.g., $n \approx 3-4$ for writ-ten English. It is obvious that in this case there are much longer correlations (orthographic, syntactic, semantic) which cannot be taken into account in this way, and which thus lead to overestimation of $h$ if sufficient care is used (with less care one is likely to obtain underestimations, as discussed below).

Improved methods take into account that one is only interested in those long blocks which have high probability. Thus also correlations should be considered only selectively, depending on their importance. The best known such methods are those based on Lempel-Ziv coding.[10,11] Here the string is coded explicitly by breaking it into non-overlapping ''words,'' and the length of the code for specifying this string of words is an upper bound for $Nh$. Technically, this is done most efficiently by preparing a ''dictionary'' of words in the form of a prefix tree.[12,13] This is a rooted tree in which each relevant word is attached to a leaf in such a way that the branch common to any two leaves corresponds just to their longest common prefix (a word $y$ is a prefix of another word $x$ if $x$ can be obtained by concatenating one or more letters to $y$). An estimator of $h$ based on similar ideas and using similar trees (but not giving upper bounds) was studied in Ref. 13 (see also Ref. 14).

A final class of methods is based on what is sometimes called ''gambling,''[8,15,16] since it is related to a method for earning maximal long time profit in a game where one has $d$ different options at any time step. The capital placed on option $i$ is multiplied by $d$ if this option is indeed realized, while the money placed on the other options is lost. It can be shown that the optimal strategy consists in sharing the total capital $K_t$ at time $t$ among all options $a$ according to their probability $p(a)$. In this case, the expected gain depends on the entropy: it is a factor $d/e^h$ per time step.

To see that the above is not just a superficial coincidence but is indeed the basis of efficient algorithms for estimating $h$ we have to make one step back. While Shannon theory is based entirely on probabilistic concepts and deals only with *average* code lengths, modern literature on information theory is mostly concerned with *individual* sequences and estimating their shortest codes. It appears from this literature that the main shortcoming of Shannon theory is that it does not take into account the information needed to describe the probability distribution itself. For the case envisaged by Shannon, namely the transmission of very long sequences with moderately complex constraints, this is irrelevant since the description of the distribution is much shorter than the description of the actual string. But in general, this is not true.

Attempts to eliminate probabilistic ideas altogether have led Kolmogorov, Chaitin, and others to algorithmic information theory.[17,18] A more practical point of view (but based on essentially the same ideas) is endorsed in Rissanen's minimum description length (MDL) principle:[19] a ''good'' encoding of a string should be one which minimizes the *total* code length. If this is applied to a string which itself is a description of some physical phenomenon, this corresponds essentially to *Occam's razor:* a good theory is a short theory.

The MDL principle was applied to entropy estimation and coding by Rissanen,[20] who called the resulting method the ''context algorithm.'' This will be discussed in detail in later sections. Here we will just point out that one need not be as radical as Chaitin and Rissanen, and can merge these ideas very fruitfully with probabilistic ideas. The best reference to algorithms following this spirit is Ref. 21. Indeed, such algorithms are implemented in most modern text compression routines.

Technically, these methods will be very similar to Rissanen's. In particular, we have to make forecasts $\hat{p}(a)$ of the probabilities $p(a)$ that $s_t = a$ under the *condition* that the previous symbols (the ''context'') had been $\ldots s_{t-2}, s_{t-1}$. This is most conveniently done by means of *suffix* trees similar to the prefix trees used in Lempel-Ziv type algorithms.

In Secs. II and III we treat more technically some concepts of information theory which will be discussed in the following, and alternative methods for estimating $h$. The concept of gambling and the associated trees will be treated in Sec. IV. Actual ansatzes for estimating $p(a)$ will be discussed in Sec. V, while applications will be presented in Sec. VI.

In the first applications, we treat symbol sequences generated by chaotic dynamical systems (logistic map and Ikeda map) in different chaotic regimes and a 1-D cellular automaton (rule 150 in Wolfram's notation[22]). We compare to alternative methods, in particular to block counting and Lempel-Ziv-like schemes.

Finally we estimate the entropies of samples of written English texts. By extrapolation to infinite text lengths, we obtain entropies of $\approx 1.5$ bits per letter. This is consistent with Shannon's results,[6] but it implies that optimal text compression algorithms should yield much higher compression rates than presently available commercial packages. We should also point out that other analyses using subjective methods like that used by Shannon tend to give even lower entropies.

## II. BLOCK ENTROPIES

We consider one-sided infinite sequences $s_1, s_2, \ldots$ where $s_t \in \{0, 1, \ldots, d-1\}$. In most examples we shall deal with $d = 2$, but everything holds also for $d > 2$ with minor modifications. We assume that these are realizations of a stochastic process $\mathbf{s}_1, \mathbf{s}_2, \ldots$ with probabilities

$$p_t(s_1, \ldots, s_n) = \text{prob}\{\mathbf{s}_{t+1} = s_1, \ldots, \mathbf{s}_{t+n} = s_n\}. \quad (1)$$

Usually it is assumed that these probabilities are stationary. In this case we can drop the index $t$ on $p_t(s_1, \ldots, s_n)$, and define *block entropies*,

$$H_n = -\sum_{s_1, \ldots, s_n} p(s_1, \ldots, s_n) \log p(s_1, \ldots, s_n). \quad (2)$$

They measure the average amount of information contained in a word of length $n$. The differential entropies,

$$h_n = H_n - H_{n-1}$$

$$= -\sum_{s_1,\ldots,s_n} p(s_1,\ldots,s_n) \log p(s_n|s_1,\ldots,s_{n-1}), \quad (3)$$

give the new information of the $n$-th symbol if the preceding $(n-1)$ symbols are known. Here, $p(s_n|s_1,\ldots,s_{n-1})$ is the conditional probability for $\mathbf{s}_n$ being $s_n$, conditioned on the previous symbols $s_1,\ldots,s_{n-1}$. The Shannon entropy is[23]

$$h = \lim_{n\to\infty} h_n. \quad (4)$$

It measures the average amount of information per symbol if all correlations and constraints are taken into account. This limit is approached monotonically from above, i.e. all $h_n$ are upper bounds on $h$.

For the numerical estimation of $h$ from a finite sequence of length $N$ one usually estimates all word probabilities $p(s_1,\ldots,s_n)$ up to some fixed $n$ by the standard likelihood estimate,

$$\hat{p}(s_1,\ldots,s_n) = \frac{n_{s_1\ldots s_n}}{N}, \quad (5)$$

where $n_{s_1\ldots s_n}$ is the number of occurrences of the word $s_1,\ldots,s_n$. (Strictly, the denominator should be $N-n+1$, but this difference is in general negligible.) From these one computes estimates $\hat{H}_n$ by inserting them into Eq. (2). Finally, an estimator $\hat{h}$ is obtained either by computing $\hat{h}_n$ and extrapolating, or simply as $\hat{h} = \lim_{n\to\infty} \hat{H}_n/n$. The latter is less influenced by statistical errors but shows slower convergence.

In practice [unless $N$ is very large and the limit in Eq. (4) is reached fast] one is confronted with serious difficulties because the number of different possible words of length $n$ increases exponentially with $n$ and so does the necessary minimum length $N$ of the sample sequence if one wants to determine $p(s_1,\ldots,s_n)$ faithfully. This means that the probability estimates $\hat{p}(s_1,\ldots,s_n)$ from which the entropies $\hat{H}_n$ are determined undergo strong fluctuations already for moderate block lengths $n$. As a consequence, the estimates $\hat{H}_n$ are usually underestimated. Formally this can be understood by looking at the expected value of $\hat{H}_n$. By use of the Kullback-Leibler inequality[24] this leads to

$$\langle \hat{H}_n \rangle = \left\langle -\sum_{s_1,\ldots,s_n} \frac{n_{s_1,\ldots,s_n}}{N} \log \frac{n_{s_1,\ldots,s_n}}{N} \right\rangle$$

$$\leq \left\langle -\sum_{s_1,\ldots,s_n} \frac{n_{s_1,\ldots,s_n}}{N} \log p(s_1,\ldots,s_n) \right\rangle$$

$$= -\sum_{s_1,\ldots,s_n} \frac{\langle n_{s_1,\ldots,s_n}\rangle}{N} \log p(s_1,\ldots,s_n) = H_n. \quad (6)$$

As long as fluctuations exist there will be a systematic underestimation of $H_n$.

A detailed computation of the expectation value of $\hat{H}_n$ up to second order in $N$ was given by Harris[25] and reads

$$\langle \hat{H}_n \rangle = H_n - \frac{M-1}{2N} + \frac{1}{12N^2}$$

$$\times \left(1 - \sum_{p(s_1,\ldots,s_n)>0} \frac{1}{p(s_1,\ldots,s_n)}\right) + \mathcal{O}(N^{-3}), \quad (7)$$

where $M$ is the number of blocks $(s_1,\ldots,s_n)$ with $p(s_1,\ldots,s_n)>0$. It is straightforward to correct for the leading $\mathcal{O}(1/N)$ bias in $\hat{H}_n$ (the second term on the rhs), as the number of different observed words is usually a good estimator for $M$. This $\mathcal{O}(1/N)$ correction term was also found independently in Refs. 26, 27. The term of order $1/N^2$ involves the unknown probabilities $p(s_1,\ldots,s_n)$, and can not be estimated reliably in general. In particular, it would not be sufficient to replace them by $\hat{p}(s_1,\ldots,s_n)$ in this term.

An alternative approach where only observables appear in the correction terms was attempted in Ref. 28. There it was assumed that each $n_{s_1,\ldots,s_n}$ is itself a random variable which should follow a Poisson distribution if $p(s_1,\ldots,s_n) \ll 1$. This leads to an asymptotic series where higher order terms become useful only for increasingly large $N$. The entropy estimate based on this assumptions [Eq. (13) of Ref. 28, corrected by adding a factor $1/n_j$ in the last term] is

$$\hat{H}_n = \sum_{j=1}^{M} \frac{n_j}{N}\left(\log N - \psi(n_j) - \frac{1}{n_j}\frac{(-1)^{n_j}}{n_j+1}\right). \quad (8)$$

Here the index $j$ counts the blocks $(s_1,\ldots,s_n)$ for which $n_{s_1,\ldots,s_n}>0$, and $\psi(x)$ is the logarithmic derivative of the gamma function. One easily sees that the leading $\mathcal{O}(1/N)$ correction is the same as in Eq. (7).

We applied the estimate Eq. (8) and the naive estimate based on Eqs. (3) and (5) to the Hénon map with standard parameters, $x_{n+1} = 1 + 0.3x_{n-1} - 1.4x_n^2$. To convert this into a bit sequence, a binary partition was used as in Ref. 5. The results are compared to the Lyapunov exponent determined by iterating the dynamics. (From Pesin's identity,[29] we know that the positive Lyapunov exponent of the Hénon map coincides with the entropy.) In Fig.1 we see that the convergence of the truncated entropy $\hat{h}_n = \hat{H}_n - \hat{H}_{n-1}$ is faster than the more conservative estimate $\hat{H}_n/n$. For long block lengths $n$, the underestimation becomes significant for small data sets, even if Eq. (8) is used.

As a second application, we took a concatenation of *ca.* 100 English texts (including the LOB corpus and texts edited by the Gutenberg project such as the Bible, Shakespeare's collective works, Moby Dick, etc.[30]) of altogether $\approx 7\times 10^7$ characters. We present in Table I estimates of $h_n$ for $n=1$ to 6 (in order to overcome the formidable storage problems, we used the trick described in Ref. 31). In the first two columns we give results for the original texts coded in 7-bit ASCII. For the last two columns, the text was converted into an alphabet of 27 characters (26 letters + blank) by converting all letters into lower case; changing all punctation marks, carriage returns and line feeds into blanks; and finally replacing all strings of consecutive blanks by single blanks. This was done in order to compare our results with previous estimates, most of which were done that way.[21] Columns 1 and 4 were obtained with the naive likelihood estimator Eq. (5), columns 2 and 5 with Eq. (8). For $n \geq 5$ we see a non-negligible difference between these two methods, indicating that this method would be rather unreliable for $n=7$, unless much larger data sets were used. We also tried to extrapolate the estimates observed for shorter texts to infinite text lengths. Within the uncertainty, these estimates (given in columns 3 and 6 of Table I) coincide for Eqs. (5) and (8). We should mention that the only previous non-subjective estimators for $h_n$ with $n \geq 4$ of written English were given in Ref. 32. They are much higher than ours, and considered as not reliable.[21]

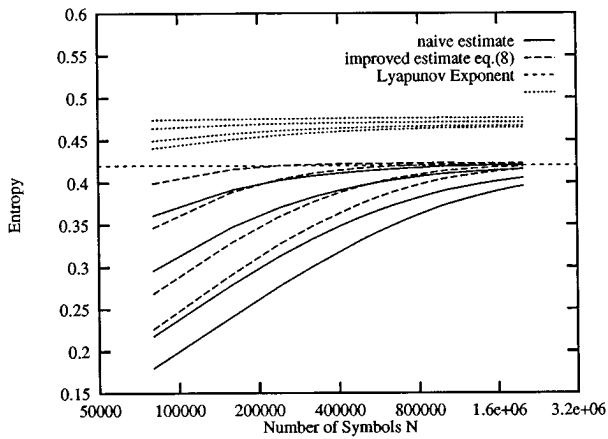

FIG. 1. Entropy estimates $\hat{H}_n/n$ (dotted lines) and $\hat{H}_n - \hat{H}_{n-1}$ (dashed lines) based on Eq. (8), for the Hénon map ($a=1.4$ and $b=0.3$) in dependence of the length $N$ of the data string. The solid lines show estimates of $H_n - H_{n-1}$ based on Eqs. (2) and (5). The Lyapunov exponent (horizontal line) is determined by numerical iteration of the tangent map. Block lengths are $n=20, 22, 24$, and 25 (top to bottom).

The corrections to the naive block entropy estimators implied by Eq. (8) are useful only if the non-zero frequencies $n_{s_1,\ldots,s_n}$ are $\gg 1$ in average, i.e. if $N \gg M$. Estimators which should work also for $N \ll M$ were proposed in several papers by Ebeling et al.[33] There, explicit assumptions are made on those $p(s_1,\ldots,s_n)$ which are too small to be estimated through their occurrences. Although these assumptions seem motivated by the McMillan theorem, we believe that this method is not very reliable as there exist no ways to check the assumptions and the results depend crucially on them. A similar but presumably safer approach was proposed in Ref. 34 where it was found that simply neglecting the information in the small $n_{s_1,\ldots,s_n}$ leads to surprisingly robust results.

The fact that $\hat{H}_n$ underestimates $H_n$ is related to the fact that we neglected the information needed to specify the probability distribution, but this relationship is not straightforward. If the distribution is simply given in form of the integers $n_{s_1,\ldots,s_n}$, the required code length can be estimated as $\leq d^n \log N$ where we have used the fact that all $n_{s_1,\ldots,s_n}$ are $\leq N$. Including this in the total information will give a safe upper bound on $H_n$, but this bound will be rather poor in general, since we did not make any effort to encode the probability distribution efficiently. Thus we will not follow this line of arguments any further. Instead, we shall discuss in the following sections alternative methods which also give upper bounds on $h$ since they essentially give the complete information needed for unique codings.

Finally we should point out that we could also replace the likelihood estimator $\hat{p}$ by some other estimator. A natural candidate would seem to be Laplace's successor rule,[35]

$$\hat{p}(s_1,\ldots,s_n) = \frac{n_{s_1\ldots s_n}+1}{N+d^n}. \quad (9)$$

Inserting this into Eq. (2) gives always a larger estimate of $H_n$ than does the likelihood estimate, Eq. (5). But it is not always closer to the correct value, in particular if the true probabilities are very far from equipartition. Just like Eq. (5), also Eq. (9) in general gives a biased estimator of $H_n$, i.e. $\langle \hat{H}_n \rangle \neq H_n$. A better strategy seems to consist in using Eq. (3), with $p(s_n|s_1,\ldots,s_{n-1})$ replaced by the Laplace estimator,

$$\hat{p}(s_n|s_1,\ldots,s_{n-1}) = \frac{n_{s_1,\ldots,s_n}+1}{n_{s_1,\ldots,s_{n-1}}+d}, \quad (10)$$

but keeping the likelihood estimator for $p(s_1,\ldots,s_n)$. Unfortunately, also this is not unbiased, and moreover it does not give definite upper or lower bounds on $h_n$.

If we have a good prior estimate for the (distribution of the) $p(s_1,\ldots,s_n)$, we can use the Bayesian estimate of $h_n$ derived in Ref. 36. But notice that also this can lead to systematic errors in either direction, if bad prior estimates are used.

## III. ZIV-LEMPEL AND SIMILAR METHODS

The most elegant upper estimates on $h$ based on explicit codings were introduced by Ziv and Lempel.[11] There, the sequence $\{s_k\}_{k=1}^N$ is broken into words $w_1, w_2, \ldots$ such that $w_1 = s_1$, and $w_{k+1}$ is the shortest new word immediately following $w_k$. For instance, the sequence $S = 110101001111\ldots$ is broken into (1)(10)(101)(0)(01) $\times(11)(1\ldots$ . In this way, each word $w_k$ with $k>1$ is an extension of some $w_j$ with $j<k$ by one single symbol $s'$. It is encoded simply by the pair $(j,s')$. The encoder and the decoder can both build the same dictionary of words, and thus the decoder can always find the new word when given $j$ and $s'$.

TABLE I. Block entropy estimates $\hat{h}_n$ in bits per character for written English, as estimated from a concatenation of several long texts of altogether $\approx 7 \times 10^7$ characters. See the text for details.

| | 7-bit ASCII | | | 27 characters | | |
|---|---|---|---|---|---|---|
| $n$ | Eq. (5) | Eq. (8) | $N \to \infty$ | Eq. (5) | Eq. (8) | $N \to \infty$ |
| 1 | 4.503 | 4.503 | 4.503 | 4.075 | 4.075 | 4.075 |
| 2 | 3.537 | 3.537 | 3.537 | 3.316 | 3.316 | 3.316 |
| 3 | 2.883 | 2.884 | 2.884 | 2.734 | 2.734 | 2.734 |
| 4 | 2.364 | 2.367 | 2.369 | 2.256 | 2.257 | 2.257 |
| 5 | 2.026 | 2.037 | 2.043 | 1.944 | 1.947 | 1.949 |
| 6 | 1.815 | 1.842 | 1.860 | 1.762 | 1.773 | 1.781 |

This encoding is efficient because for sequences of low entropy there are strong repetitions, such that the average length of the code words $w_k$ increases faster, and the number of needed pairs $(j,s')$ slower, than for high entropy sequences. More precisely, the expected word length $L(w)$ increases with the number of symbols $N$ roughly like

$$\langle L(w)\rangle \approx \frac{\log N}{h}, \tag{11}$$

where the logarithm is taken with base $d$, the cardinality of the alphabet. Since the information needed to encode a pair $(j,s')$ increases like $\log N$, the information per symbol is $\approx h$.

This Ziv-Lempel (ZL) coding is indeed a simplification of an earlier algorithm by Lempel and Ziv,[10] called LZ coding in the following. There $S$ is also broken up into a chain of words $w_1 w_2 \ldots$, but a word $w_k$ is not necessarily an extension of a previous word $w_j$. Instead it can be an extension of any substring of $S$ which starts before $w_k$ (and maybe overlaps with it).

In the above example we get the different parsing $(1) \times (10)(10100)(111)(1\ldots$ . This is obviously more efficient than the ZL parsing in the sense that the average word length increases faster and the algorithm can make better use of long range correlations. But the coding/decoding procedure is now more complicated, whence it is not much used in practice. Also, the code length per word is slightly larger, so that it is not clear whether its compression performance is indeed superior to ZL for small $N$. In any case, the convergence of the compression rate with $N$ is not well understood theoretically in either of these schemes.

For a more formal description we first need some more terminology. We call $s_i^j$ the substring of $S$ starting at position $i$ and ending at $j$, $s_i^j = s_i,\ldots,s_j$. If $i=1$, i.e. for the prefix of $S$ of length $j$, we write $s^j$. Let the word $w_k$ start at position $j_k$, i.e. $w_k = s_{j_k}^{j_k+L-1}$ for some $L>0$. Its length $L$ is determined as

$$L(w_k) = \min\{p:s_{j_k}^{j_k+p-1} \neq s_m^{m+p-1}, \quad 1 \leq m < j_k\}. \tag{12}$$

Then for both, the LZ and the ZL schemes, the entropy of stationary ergodic sources is

$$h = \lim_{N\to\infty} \frac{\log N}{\langle L(w)\rangle}. \tag{13}$$

Here $L(w)$, the length of the word $w$, is essentially (up to a constant) just the length of maximal repeats: of entire previous words in the case of ZL, and of arbitrary previously seen strings for LZ.

We now ask whether we can generalize to arbitrary repeats. Let us define for each $i$,

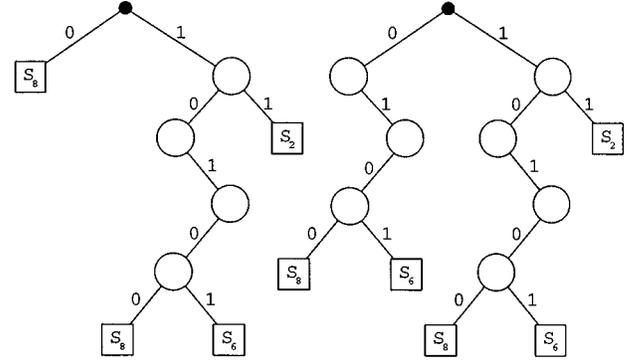

FIG. 2. Prefix tree $T(4)$ of the string $110101001111\ldots$ up to $n=4$ (left), and the tree $T(5)$ (right). The set of strings assigned to the leaves of $T(5)$ are $\{11,10101,0101,10100,0100\}$. The $S_i$ are pointers to the positions $i$ where the substrings corresponding to the respective leaves end.

$$L^i = \min\{p:s_i^{i+p-1} \neq s_j^{j+p-1}, \quad 1 \leq j < i\}, \tag{14}$$

i.e. $L^i$ is the length of the shortest prefix of $s_i, s_{i+1}, \ldots$ which is not a prefix of any other $s_j, s_{j+1}, \ldots$ with $j<i$. Then it can be shown that the estimator,

$$\hat{h}_N = \frac{N \log N}{\sum_{i=1}^N L^i}, \tag{15}$$

converges to $h$ if the source is a stationary Markov chain of finite order.[14] Unfortunately this is not necessarily correct if the source is just ergodic but not Markov. A similar estimator to (15) is that used in Ref. 13, and the same remarks hold also for that. Thus it is not clear whether $\hat{h}_N$ or the estimator used in Ref. 13 can be applied to cases like written English, where it seems to give definitely smaller estimates than either LZ or ZL.

The optimal data structure for the above estimation procedures is a so-called prefix tree construction.[13] For its description we consider rooted trees which consist of branching (or internal) nodes, and of leaves (external nodes) that store *keys* (Fig. 2). The meaning of the keys will be explained below. Internal nodes are drawn as circles and the leaves as squares. The dot at the top denotes the root of the tree.

To each node, one assigns a substring by following the path from the root to this node. For any $i$, $1 \leq i \leq N$, we define $w_i^N$ as the shortest prefix of $s_i, s_{i+1}, \ldots$ which is not a prefix of any other $s_j, s_{j+1}, \ldots$ with $j \leq N$ and $j \neq i$. By definition it follows that for fixed $N$, all $w_i^N$ are distinct. The minimal prefix tree of the string $s = s^N$ is the tree which represents all words $\{w_i^N : 1 \leq i \leq N\}$ as paths from the root to the leaves. In this way, each leaf of the tree corresponds to a string $s_i, s_{i+1}, \ldots$, for $1 \leq i \leq N$.

The construction of the tree is done recursively. For $N=0$, i.e. before any symbol has been received, we start with the empty tree $T(0)$ consisting only of the root. Suppose the tree $T(N)$ represents the string $s^N$. One obtains the tree $T(N+1)$ adding a leaf as follows. One starts from the root of $T(N)$ and takes right or left branches corresponding to the observed symbols $s_{N+1}, s_{N+2}, \ldots$ . This process terminates when the path of the actual prefix differs from the path already in the tree, or when a leaf is reached (remember that we assumed the sequence to be one-sided infinite!).

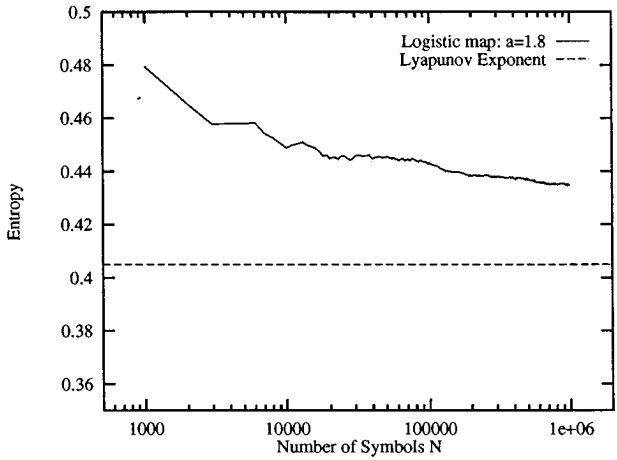

FIG. 3. Convergence of the Lempel-Ziv like estimate $\hat{h}_N$ of Eq. (15) for the logistic map in a typical chaotic regime $a=1.8$ in dependence of the length $N$ of the string.

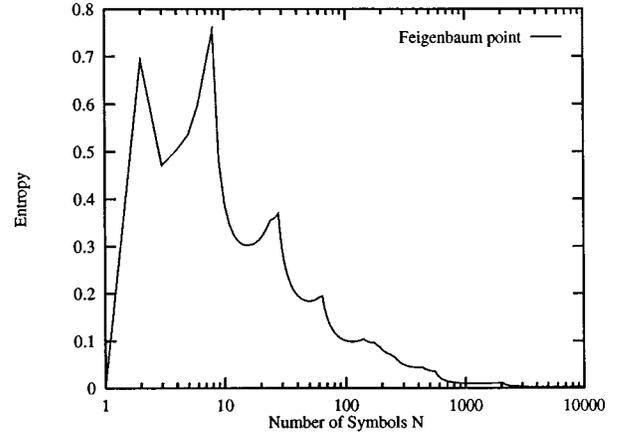

FIG. 4. Convergence of the estimate $\hat{h}_N$ for the logistic map at the Feigenbaum point in dependence of the length $N$ of the string. The cusps indicate the beginnings of long repetitions in the data string.

In the first case a new leaf is generated and the pointer (key) to the position $N+L^{N+1}$ is stored in it. [For the definition of $L^N$ see Eq. (12).] If we reach a leaf before termination, we replace this leaf by an internal node. Using the position in the string stored in the old leaf, we follow both substrings forward. We add further internal nodes as long as they coincide, and add leaves when they first disagree. Finally we store both keys in these leaves. In Fig. 2 we see the prefix trees $T(4)$ and $T(5)$ for the example 110101001111... .

We have applied Eq. (15) to the logistic map $x_{n+1}=1-ax_n^2$ with a binary partition at the critical point $x=0$. There a suitable symbolic dynamics is defined as $s=0$ if $x<0$ and $s=1$ for $x>0$. Interesting cases are the typical non-trivial chaotic range represented by $a=1.8$ (Fig. 3), and at the Feigenbaum point $a=1.40115518...$[37] where the entropy is zero, but the block entropies diverge logarithmically (Fig. 4).

## IV. "GAMBLING" AND SUFFIX TREES

Another class of entropy estimates is related to *gambling*.[15,16] Let $s_1,...,s_t$ be again a sequence of random variables taking values in a finite set $\{0,...,d-1\}$, and suppose a gambler is allowed to bet on these random variables according to his individual strategy. Knowing the past outcomes $s_1,...,s_{t-1}$, he distributes his limited wealth at the beginning of every round $t$ over the possible outcomes of $s_t$. The gambler will collect the return of his investment at the end of each round when the outcome $s_t$ is revealed, and start the next round. The return of the bets at the end of the round $t$ is $d$ times the amount that was invested in the actual outcome $s_t$. If the gambler starts with initial wealth $K_0=1$ and places the fraction $q(s_t|s^{t-1})$ on every possible outcome $s_t$ during round $t$, then his wealth after $n$ rounds amounts to

$$K_n = d^n q(s^n), \quad (16)$$

where

$$q(s^n) = \prod_{t=1}^{n} q(s_t|s^{t-1}). \quad (17)$$

In order to maximize his long term wealth, the gambler should not maximize the expectation value of $K_{t+1}/K_t$, but that of $\log(K_{t+1}/K_t)$. When using the strategy $q(s_t|s^{t-1})$, this is

$$\left\langle \log \frac{K_{t+1}}{K_t} \right\rangle = \log d + \langle \log q(s_{t+1}|s^t) \rangle. \quad (18)$$

If the gambler knows the true distribution of the process, to maximize the expected grow rate $\log(K_{t+1}/K_t)$ of his wealth, he thus should place the bets at the beginning of round $t$ proportional to the conditional probabilities $p(s_t|s^{t-1})$ on every outcome $s_t$.[16] In the case of stationary ergodic processes, the expected logarithmic maximum growth rate of the capital is then equal to

$$\left\langle \log \frac{K_n}{K_{n-1}} \right\rangle = \log d - h_n. \quad (19)$$

An interesting question is how to gamble on a stationary sequence whose distribution is unknown and has to be estimated on the basis of the past sequence $s_1,...,s_{t-1}$. Formally, a gambling scheme $q(s_t|s^{t-1})$ is called *universal* if for any stationary ergodic process

$$\lim_{n\to\infty} \frac{1}{n} \log K_n = \log d - h, \quad (20)$$

with probability 1.

For binary sequences Cover[15] exhibited a scheme for gambling with growth $K_n \geq 2^{n-C(s^n)}$. Here $C(s^n)$ denotes the length of the shortest binary program on a Turing ma-chine which computes $s^n$, when no program is allowed to be the prefix of any other [$C(s^n)$ is also called *algorithmic complexity* of string $s^n$[18]]. Thus, the growth rate of the gambler's wealth is related to the ability of the gambler to compress the sequence from $n$ to $C(s^n)$ bits so that he can double his wealth at least $n-C$ times. Unfortunately this scheme is not *computable* because it requires the evaluation of the complexity $C$.

Alternatively, gambling schemes can be used for the purpose of data compression via arithmetic coding.[21,38] For any set of probabilities $q(s_t|s^{t-1})$ used in arithmetic coding, the average code length — which is an upper bound on the entropy — is given by $\approx -\log q(s^n)$ for data strings of length $n$. Notice that we do not need necessarily either stationarity or ergodicity. But we do need the existence of a probability measure $p$. If this is non-stationary, we do not have any guarantee that any entropy estimate converges, and if it is non-ergodic, we measure just the entropy of the particular sequence. Again, the main problem in practical applications consists in estimating this $p$, since optimality is reached when $q=p$. The construction of "universal" estimators $\hat{p}$ for which the average code length actually converges to the entropy $h$ is in general an unsolved problem.

The simplest class of models for which universal coding schemes can be found are Markov chains of known order. Let us call, following Ref. 20, the conditioning string $s_{t-r}^{t-1}$ in $p(s_t|s_{t-r}^{t-1})$ as the *context* of order $r$ in which $s_t$ appears. A Markov chain of order $r$ is defined such that all contexts of order $>r$ are irrelevant since $p(s_t|s_{t-r'}^{t-1})=p(s_t|s_{t-r}^{t-1})$ for all $r'>r$. Here, the Laplace estimator, Eq. (9), with $n=r$ is universal. But in practice $r$ can become very large. Since the number of relevant blocks increase exponentially with $r$, using Laplace's estimator becomes unfeasible in this case. The situation is even worse if the order $r$ is not known.

A less trivial case is that of "finite state sources." These are sequences emitted by Markovian (i.e., finite memory) sources. But in contrast to Markov chains, the state of the "memory" is not seen from the context but is hidden to the observer. A special subclass for which efficient universal coding schemes have been proven in Refs. 20, 41, and 42 are "finite memory tree sources." Technically, these are Markov chains, but they are of such high order that a straightforward approach using contexts of fixed order is practically useless. Instead, here the order of "relevant" contexts depend strongly on the symbol to be forecasted. As we shall see, this can substantially reduce the complexity of the model. Heuristically, similar algorithms had been used in Ref. 21 and in the papers quoted there. We will not give here a formal definition of finite memory tree sources, since we do not have any proof that our applications are of this type (all indications are to the contrary). Nevertheless, we shall see that the concept of using context of variable length is extremely useful.

Algorithms for estimating $p(s_t|s_{t-r}^{t-1})$ will be given in the next section. Irrespective of any specific algorithm for the estimation of $p(s_t|s^{t-1})$ or of $p(s_t|s_{t-r}^{t-1})$, this means that we have to prepare the data such that we have easy and fast access to previous occurrences of variable length contexts, and to the symbols followed them. The natural construction for a complete count of the repetitions in the data string is a suffix tree. It is similar to the prefix tree in the previous section, but is read in the reversed direction. Therefore, instead of moving forward in the string as in constructing the prefix tree, one moves backward into the past. Correspondingly, we assume that the string is left-sided infinite. In practice we can achieve this by attaching an arbitrary but fixed aperiodic precursor (e.g., the binary digits of $1/\sqrt{2}$ in reversed order, ...01101), indicated in the following by a semicolon. Moreover we shall assume for simplicity $d=2$, though all constructions work for arbitrary $d$. In the left of Fig. 5 we see the tree $T(4)$ for the last $N=4$ bits of the string $s^4=\ldots 1;1010$. In contrast to the prefix trees, here each of the $N$ leaves corresponds to a *suffix* of the string which end at position $k$, $k=1,\ldots,N$. At each leaf is stored a pointer to the start positions of its suffix. The tree $T(5)$ of the string ...1;10101 extended by one symbol is shown in the right of Fig. 5.

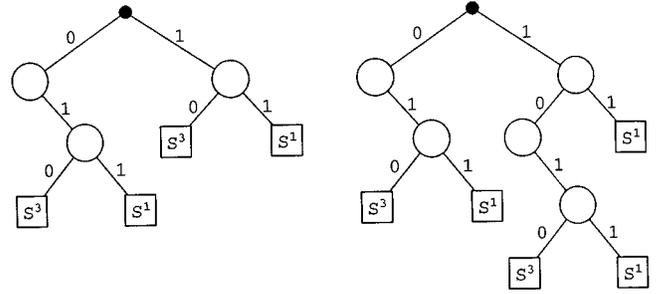

FIG. 5. (Left) Suffix tree for the last 4 bits of the binary string ...1;1010 with suffixes {11,110,01,010}. (Right) Tree after extending the string by one symbol to ...1;10101.

Finally, in each internal node we store how many times it was visited during the construction of the tree. The contexts for $s_t=0$, $1\leq t\leq N$, are paths on the left part of the tree (with the numbers in the internal nodes indicating their frequencies), while the contexts for $s_t=1$ are *parallel* paths on the right half of the tree. For an estimate of $p(s_{N+1}|s^N)$ we thus want to access pairs of paths which are parallel except for the last step. The frequencies in the left half of the tree are those of contexts which had preceded $s=0$, while those on the right half had been followed by $s=1$ (Fig. 6).

By using the conditional frequencies read off the tree $T(N)$, one can thus determine the estimates $\hat{p}(s_{N+1}|s^N)$ for the next outcome to be $s_{N+1}$. Actually, the needed information can be stored most efficiently in a tree which is a superposition of the left and the right branches of the above tree (Fig. 7). Thus, each node of the modified tree is associated with a set of $d$ symbol counts, where $d$ is the cardinality of the alphabet.

In general a tree constructed in this way grows very fast with the length of the data string and therefore also reaches very fast the limits of the memory resources. A tree which does not grow as fast, called *Rissanen's suffix tree* in the following, is used in Ref. 43. It works as follows.

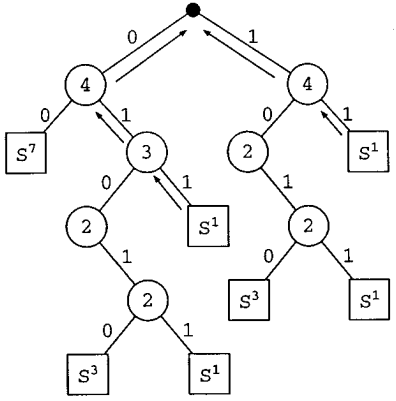

FIG. 6. Suffix tree of the binary string …11010100. The internal nodes contain conditional frequencies. Looking at the arrows in the left branch, substring 11 is followed one times by 0, and symbol 1 is followed three times by 0. In the right branch, 1 is followed one times by 1, and occurs four times in all.

We start with the root, with its symbol counts set to zero. Recursively, having constructed the tree $T(t)$ from $s^t$, read the symbol $s_{t+1}$. Climb the tree according to the path defined by $s_t, s_{t-1}, \ldots$, and increment the count of symbol $s_{t+1}$ by one for every node visited until the deepest node, say $s_t, s_{t-1}, \ldots, s_{t-j+1}$, is reached. If the last update count becomes at least 2, create a single new node $s_t, s_{t-1}, \ldots, s_{t-j}$, and initialize its symbol counts to zero, except for the symbol $s_{t+1}$, whose count is set to 1.

Although Rissanen's tree does *not* count all repetitions of the contexts, it grows where repeated patterns occur frequently. In the case of the suffix tree all repetitions of contexts and their corresponding counts are collected. But on the other hand the suffix tree grows faster than Rissanen's tree and is thus not suitable for very long sequences or sequences with extremely strong long range correlations. We should point out that the memory needed to store either tree increases linearly with the length $N$ of the symbol sequence (for $h > 0$), whence available workspace often is a limiting factor. Also, Rissanen's tree mainly leaves out contexts which had occurred in the distant past. For non-stationary sequences this might be advantageous since such contexts might lead to wrong estimates.

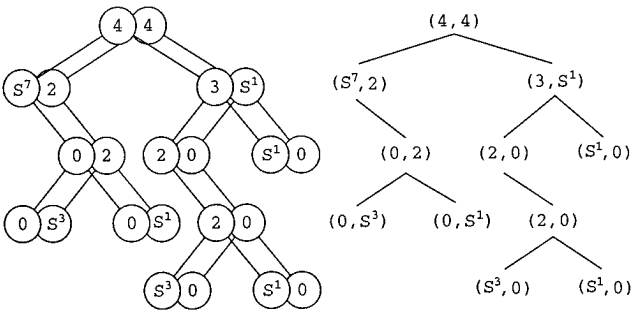

FIG. 7. Superposition of the subtrees representing the string 11010100. For simplification, all nodes are expressed by circles (left); compact version of the superposed tree (right).

## V. MODELING PROBABILITIES

In this section, we will present different strategies which have been proposed for actually estimating $p$. There exists no globally optimal strategy. As in the case of block entropies discussed in Sec. II, even an estimator which is "optimal" in being unbiased ($\langle \hat{p} \rangle = p$) does not give an unbiased entropy estimator. But we shall arrive at one strategy which we have found to be close to optimal in most cases.

### A. Bayesian probability estimation

Let us assume that we want to estimate the probability for the $t$-th symbol $s_t$, based on its contexts $s_{t-j}^{t-1}$ of length $j = 0, 1, \ldots$. Suppressing in the following the index $t$, we denote by $\sigma_j$ the context $s_{t-j}^{t-1}$ and by $n_j$ the number of times this context had appeared previously in the string, at times $t_k$, $k = 1, \ldots, n_j$. Similarly, $n_j^{(a)}$ is the number of times that $\sigma_j$ had been followed by symbol $a$, i.e. $s_{t_k} = a$. These are exactly the frequencies stored in the nodes of the suffix tree $T(t-1)$. Obviously, $n_j = \Sigma_a n_j^{(a)}$.

A first guess for $\hat{p}(s_t = a | \sigma_j)$ could be the likelihood estimator $n_j^{(a)}/n_j$. Unfortunately, this would lead to $\hat{p} = 0$ if $n_j^{(a)} = 0$, and consequently to a divergent code length if a symbol appears for the first time in the given context. Thus, all symbols that possibly can occur must get probability estimators greater than zero. But which probability should one assign to an event which did never occur? This so-called *zero-frequency* problem is commonly treated from a Bayesian point of view.[39] For an alphabet of $d$ symbols, a consistent (though not necessary optimal) class of estimators is given by

$$\hat{p}(s_t = a | \sigma_j) = \frac{n_j^{(a)} + \beta}{n_j + \beta d}, \quad \text{with } \beta > 0. \quad (21)$$

These include Laplace's rule which is the Bayes' estimator for the uniform prior ($\beta = 1$), and the Krichevsky-Trofimov (KT) estimate[40] which uses $\beta = 1/2$. A detailed justification of Eq. (21) can be found in Ref. 39.

Within a Bayesian approach, each context leads to a posterior probability distribution $P(\mathbf{p} | \sigma_j)$ where $\mathbf{p} = (\text{prob}(s_t = 1 | \sigma_j), \ldots, \text{prob}(s_t = d | \sigma_j))$. It is easily seen that the optimal choice for $\hat{\mathbf{p}}$ is the average value of $\mathbf{p}$ with respect to $P$, denoted by $\bar{\mathbf{p}}$ (for a uniform prior, this is just the Laplace estimator). To see this, let us assume that we have used the correct prior. Then $P(\mathbf{p} | \sigma)$ is indeed the correct posterior distribution, and the average code length for $s_t$ is $\int d^d p\, P(\mathbf{p} | \sigma) \Sigma_{i=1}^d p_i \log \hat{p}_i = -\Sigma_{i=1}^d \bar{p}_i \log \hat{p}_i$. By the Kullback-Leibler inequality this is minimal if $\hat{p}_i = \bar{p}_i$. As usual in Bayesian approaches, the efficiency of the method depends on the availability of a good prior. In the following we will only use Eq. (21), with the understanding that this is a major source of future improvements.

After having an estimator for $\hat{p}(s_t | \sigma_j)$, the next problem is how to chose the most "successful" rank $j$ which should determine the actual estimate $\hat{p}$. In general this is quite intricate. If one prefers to use long contexts (i.e., those close to the leaves in the tree), then these estimates are determined by a very small number of events. In contrast, estimates based on nodes near the root (i.e. short contexts) are in general statistically more reliable, but they only lead to lower order entropies since they do not take into account long range correlations.

In a Bayesian framework, one should indeed not choose a single $j$ for estimating $\hat{p}$. Instead, since there will also be a prior distribution for the "correct" $j$, one must expect $\hat{p}$ to be obtained from a posterior distribution over different $j$'s. In the next subsections we shall first discuss a method which is based on selecting a unique $j$, and then a (somewhat *ad hoc*) method for averaging over different $j$'s.

### B. Rissanen's method

A strategy for estimating the optimal context length has been suggested in Ref. 43. Let us denote by $z_j$ the concatenation of all symbols following previous appearances of the context $\sigma_j$,

$$z_j = s_{t_1}, s_{t_2}, \ldots s_{t_{n_j}}. \quad (22)$$

The total code length for all these symbols, using estimates based on context $\sigma_j$, is

$$l(z_j|\sigma_j) = -\sum_{i=1}^{n_j} \log \hat{p}(s_{t_i}|\sigma_j). \quad (23)$$

Any symbol which has occurred at a context $\sigma_j$ in the tree has also occurred at the context $\sigma_{j-1}$ corresponding to the parent node in the suffix tree. This suggests comparing the context models based on $\sigma_j$ and $\sigma_{j-1}$ by computing the code length difference

$$\Delta_j = l(z_j|\sigma_j) - l(z_j|\sigma_{j-1}). \quad (24)$$

We should stress that $z_j$ contains only a part of the symbols which followed after the context $\sigma_{j-1}$ since the latter is shorter and thus less specific, and $l(z_j|\sigma_{j-1})$ measures the success of $\sigma_{j-1}$ for *those occurrences only*.

When the sign of $\Delta_j$ is positive, then $\sigma_{j-1}$ has been more efficient in the past than $\sigma_j$. It is then a good guess that it will be more efficient for the present symbol as well. The most obvious strategy[43] is thus to take the shortest context (i.e., the smallest $j$) for which $\Delta_j < 0$ and $\Delta_{j+1} > 0$.

Unless the source is Markov, the node $\sigma_j$ will win against its parent node $\sigma_{j-1}$ in course of time, and $\Delta_j$ will finally become negative for $t \to \infty$. Thus finally longer and longer contexts will be chosen for long sequences. This implies the risk that the frequencies $n_j$ for the chosen contexts are small. While the above rule for selecting $j$ is certainly reliable in the limit of large $n_j$, this is not necessarily so for contexts which had occurred only rarely. To avoid selecting too long contexts one can use at least two modifications.

In the first, one defines a minimal number $n_{\min}$, and discards any context with $n_j < n_{\min}$. A better strategy is to define a threshold $\delta \geq 0$, and to choose context length $j$ only if $\Delta_j < -\delta$. Otherwise said, the selected context has the smallest $j$ for which $\Delta_j < -\delta$ and $\Delta_{j+1} > 0$. If there is no $j$ for which this is fulfilled, the root ($j=0$) is taken.

To speed up the algorithm, the code length differences $\Delta_j$ are stored in the corresponding node of the tree and are updated recursively while the tree is constructed. At the beginning, all $\Delta_j$ are set to zero, except for the root node for which $\Delta_0 = -1$.

### C. Superposition of probabilities

Indeed, it is not clear if strategies which use only single nodes for the prediction will be most successful. Let us assume that the largest context for which $n_j > 0$ has length $j = r$. Several authors suggested to use weighted averages over all context lengths $\leq r$,

$$\hat{P}_t(a) = \sum_{j=0}^{r} c_j^{(t)} \hat{p}(a|\sigma_j), \quad (25)$$

where the positive weights $c_j^{(t)}$ are properly normalized, and $\hat{P}_t(a)$ is the final estimator for the probability that $s_t = a$.

(i) In Ref. 44, it was proposed to take the weights independent of $t$, i.e. $c_j^{(t)} \equiv c_j$ for all $t$, and to minimize $h$ with respect to them. But this seems too rigid, as we should expect the strengths of correlations to depend on the context. For this reason we shall not discuss this further.

(ii) Another way to compute the weights is *adaptively* as in Ref. 21 (Sec. VI). These authors consider the estimated probabilities for encountering a symbol which had not appeared previously after the context of length $j$, $e_j = \text{prob}\{n_j^{(s_t)} = 0\}$. They argue that $c_j^{(t)}$ should be large if $e_j$ is small (i.e., if context $\sigma_j$ is statistically relevant), and if all $e_i$ with $i > j$ are large (i.e., all longer contexts are irrelevant). More precisely, they chose

$$c_j^{(t)} = (1-e_j) \prod_{i=j+1}^{r} e_i, \quad 0 \leq j < r,$$
$$c_r^{(t)} = 1 - e_r. \quad (26)$$

It is easily verified that these weights are properly normed, $\Sigma_j c_j^{(t)} = 1$. As is usual for estimated probabilities, there is no general theory for the optimal determination of the $e_j$. Of course, they should be 1 if no symbol has been observed, i.e. if $n_j = 0$. On the other hand, if there have been very many observations ($n_j \gg 1$), then the $e_j$ should approach to zero. Thus, the following ansatz was made in Ref. 21:

$$e_j = \frac{q}{n_j + q}, \quad q > 0. \quad (27)$$

In practice, entropy estimates $\hat{h}_N$ based on Eqs. (21) and (25)–(27) are remarkably good for written texts, and they are very robust against changes in the parameter $q$. But we found them to be much worse when applied to sequences produced by chaotic dynamical systems. The reason for this seems to be that this algorithm puts too much weight on long contexts. Thus we will not use them in the following.

### D. Global probability estimates

For all strategies discussed so far, we have to specify certain parameters that can not be fixed *a priori*. Even if the strategy can be shown to be universal, these parameters still can have a big influence on the convergence of the entropy estimate. An estimator for binary sources which is supposed not to be dependent on arbitrary model parameters is suggested in Ref. 42.

In contrast to the above methods, all of which make ansatzes for the conditional probabilities $p(s_t|s^{t-1})$ and compute the total probability estimator as a product $\hat{p}(s^t) = \Pi_{t'\leq t}\hat{p}(s_{t'}|s^{t'-1})$, these authors prove universality for an estimator which gives $\hat{p}(s^t)$ directly as a weighted average. They did not show that this method gives good estimates also for finite sequence lengths, and it is not clear that this method converges faster than, e.g., Rissanen's method. Thus we have not studied it further.

In the next section we will apply Rissanen's method based on Eqs. (21) and (24). Whenever it was feasible with our computational resources, we used the full suffix tree instead of Rissanen's tree. To avoid too long contexts, we use a threshold $\delta$ as discussed above. The values of $\beta$ used in Eq. (21) and of $\delta$ were chosen differently in different applications. This did not give always the smallest entropy for short sequences. But even if it did not, it never gave a value much larger than the minimum, and it improved in general quickly with sequence length. In particular, it seemed to lead in all cases to the smallest estimates for infinite sequence length, when using the extrapolation discussed in the next section.

## VI. APPLICATIONS

In all non-trivial cases we observed that $\hat{h}_N$ converged very slowly with $N$. This suggests that a careful extrapolation to $N \to \infty$ is essential when estimating $h$. Unfortunately, even in the case of known sources it seems not easy to derive the $N$-dependence of $\hat{h}_N$. This is obviously even worse for sequences with unknown statistics such as written texts or typical chaotic symbol sequences.

But we found empirically that the following ansatz yields in all cases an excellent fit:

$$\hat{h}_N \approx h + c\frac{\log N}{N^\gamma}, \quad \gamma > 0. \quad (28)$$

As we said, we can not prove this, but we verified it numerically for several very different types of sources. A similar ansatz, but for block entropy based estimates, was made in Ref. 45.

### A. Chaotic dynamical systems and cellular automata

We begin with the logistic map $x_{n+1} = 1 - ax_n^2$ with a binary generating partition at $x=0$. Interesting cases are the typical non-trivial chaotic range $a=1.8$, the case $a=1.7499$ (Fig. 8) where it is very strongly intermittent (there is a Pomeau-Manneville intermittency point[46] at $a=1.75$) and at the Feigenbaum point (Fig. 9).

In all three cases, Eq. (28) gives very good fits. They (and the fits in all cases to follow) were done as least square fits. At the Feigenbaum point, we know that $h=0$, whence Eq. (28) involves only 2 free parameters. In the two cases shown in Fig. 8, the entropy $h$ was used as a fit parameter. These fits gave $h(a=1.8) = 0.404$ and $h(a=1.7499)$ $=0.186$. Due to Pesin's identity,[29] these values should coincide with the Lyapunov exponents, which were found to be 0.405, respectively, 0.184.

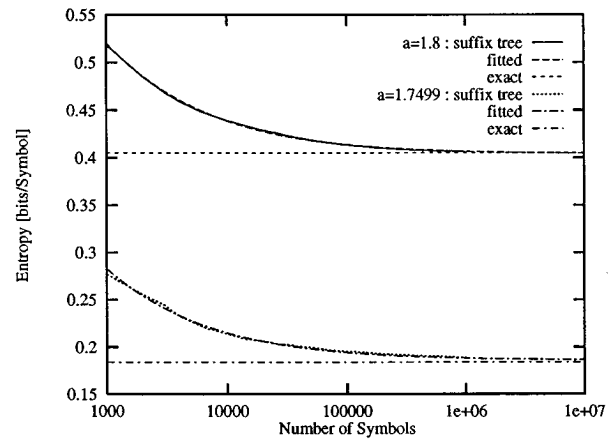

FIG. 8. Entropy estimation of the logistic map $x_{n+1} = 1 - ax_n^2$ with parameter $a=1.8$ and $a=1.7499$ by using full suffix trees. The curves are averages over 50 samples, each consisting of $N=10^6$ symbols. The straight lines are the Lyapunov exponents numerical determined by use of the dynamics. The fit with Eq. (28) coincides very well with the estimated data.

Next we applied the algorithm to a sequence generated by the Ikeda map[47] for standard parameters,

$$z_{t+1} = 1 + 0.9z_t e^{0.4i - 6i/(1+|z_t|^2)}. \quad (29)$$

Again we used a binary generating partition.[48] The convergence of the entropy estimate is shown in Fig. 10. We can again use Pesin's identity to compare the fitted value of $h$ to the positive Lyapunov exponent, with excellent agreement. Similarly good agreement is found with the $N$-dependence of Eq. (28).

A quite different type of sequence is obtained by iterating a 1-D cellular automaton (rule 150 in Wolfram's[22] notation) for a small and fixed number of times, starting from an input string with low but known entropy. Since rule 150 is bijective, the entropy of the input string is preserved at every time step. But after $T$ iterations one needs context length $>2T$ to recover the correct value of the entropy. In particular, we started with random input strings with a ratio 1:19 of 0's and 1's. This gives an entropy of 0.286 bits. The convergence of the algorithm for $T=2$, 4 and 6 iterations is shown in Fig. 11. As expected, the convergence becomes slower when increasing the number $T$ of iterations. In particular, the fitted values of $h$ were 0.286, 0.293, and 0.294 for $T=2,4,$ and 6. But the validity of the scaling law seems to be unaffected.

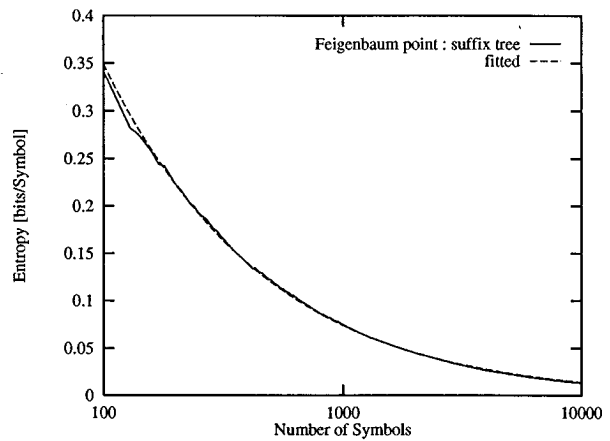

FIG. 9. Entropy estimation of the logistic map at the Feigenbaum point for $10^4$ symbols by use of a full suffix tree. The exact value of the entropy is zero.

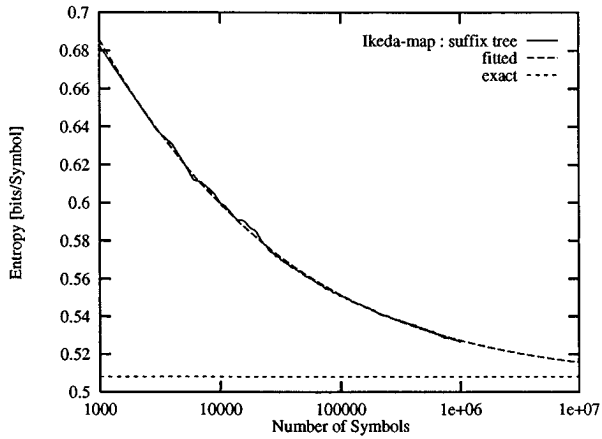

FIG. 10. Entropy estimate of the Ikeda ($N=10^6$). The Lyapunov exponent is $\approx 0.508$, the estimated asymptotic value for the entropy is 0.506. The fit describes very well the estimated data.

### B. Natural languages

These examples suggest already that the ansatz (28) works very well, independently of the particular type of sources considered. Our last tests were done for written natural languages. We will present only results for English texts.

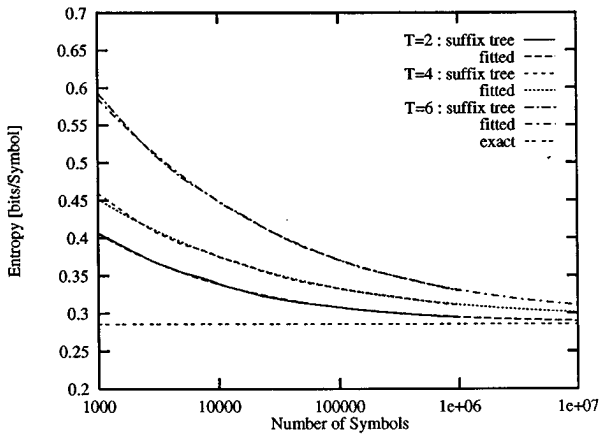

FIG. 11. Entropy of the cellular automaton rule 150 for $p_0(t=0)=0.05$ and for a different number of iterations $T=2,4,6$. Each data curve is obtained by averaging over 50 samples, each consisting of $N=10^6$ symbols.

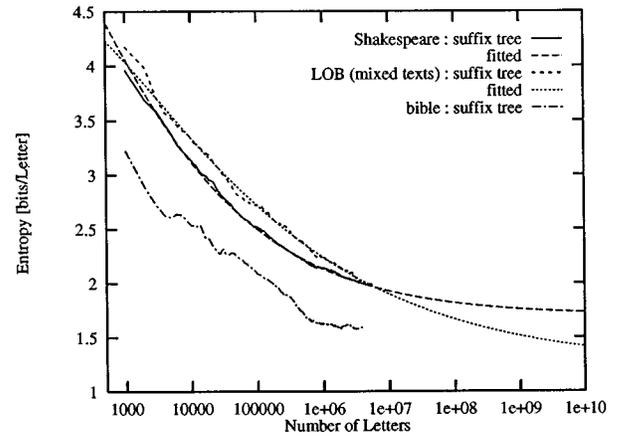

FIG. 12. Entropy estimate of Shakespeare's collected works, ($N=4791000$ letters), LOB corpus ($N=5625000$), and the Bible ($N=4014000$). All are estimated by using Rissanen's tree. For the first two texts the extrapolation yields the asymptotic entropies $\approx 1.7$ (Shakespeare) and $\approx 1.25$ bits per letter (LOB corpus). An extrapolation of the curve for the Bible seems difficult due to the large fluctuations.

Of course, any entropy estimate for natural languages is much more delicate, since it is not clear *a priori* that languages can be described probabilistically, and that concepts like stationary or ergodicity are meaningful.

Already from the work of Shannon and others[6–8] it appears that in ordinary literary English the long range statistical effects (up to hundred letters) reduce the entropy to something of the order of one bit per letter and that this might still be reduced when structure extending over paragraphs, chapters, etc. is included. Thus one can not expect a saturation of the gambling entropy estimate with present-day computer facilities. Notice that the longest structures "learned" by gambling algorithms are given by the average height of the suffix tree, and this grows only logarithmically with the text length.

The samples we consider are mainly the collected works of Shakespeare ($N\approx 4791000$ letters), the LOB corpus (mixed texts from various newspapers; $N\approx 5625000$), and the King James Bible with $N\approx 4014000$.[30] In order to be able to compare our results with those in the literature,[6,9,8] we modified the texts as described in Sec. II. Thus our alphabet $\{a,b,\ldots,z,\text{blank}\}$ consists of 27 letters. If no frequencies were taken into account its entropy would be $\log_2 27 \approx 4.76$ bits per letter. The entropy based on single letter frequencies are $\approx 4.1$ bits per letter for all considered texts.

As we have already said in Sec. V, a natural extension of the Laplace estimate, Eq. (10) is $(k+\beta)/(n+\beta d)$, where $d$ is the cardinality of the alphabet. We found numerically that the best estimates were obtained with $\beta \approx 1/d$, whence

$$\hat{p} = \frac{k+1/d}{n+1} \qquad (30)$$

yields better entropy estimates than the Laplace estimator.

For the first two samples we see that Eq. (28) again fits the data very well (Fig. 12). The asymptotic estimates of the entropies are ≈1.7 for Shakespeare and ≈1.25 for the LOB corpus. This is in good agreement with the experiments of Shannon and others,[6,9,7,8] given the spread of the latter. The curve for the third sample (Bible) is not very smooth, and an extrapolation seems not very reliable. But it seems to give an entropy similar to the other two samples.

In addition, we also studied the text called ''book1'' ($N=768771$) in Ref. 21, coded in 7 bit ASCII (i.e., with capitals and all punctuation marks) as in that reference. For this, our algorithm gave an entropy of 2.46 bits/letter without extrapolation (1.5 with extrapolation), compared to the best value of 2.48 bits/letter in Ref. 21.

The irregular behavior of the Bible is easily explained by its linguistic inhomogeneity (this was also observed as a large ''random walk exponent'' in Refs. 49 and 50). For example, compare the genealogical enumerations of Genesis **36** or Numbers **1** (abounding with proper names and with endlessly repeated nearly identical phrases) with the intricate philosophical discussions of Paul's letters. Each time the statistical features of the text changes, the old tree becomes less efficient, and the effective entropy increases. On the other hand, the very long repetitive enumerations give particularly low entropies.

The difference between the LOB corpus and Shakespeare's works is explained similarly. Although the extrapolated $h$ is lower for the LOB corpus, the LOB entropy estimates for finite $N$ ($10^3 < N < 10^6$) are higher. The latter is explained by the fact that the LOB corpus consists of short unrelated pieces of text. After each such piece the algorithm has to cope with new statistics, whence the entropy $\hat{h}_N$ is large on intermediate values of $N$. But ultimately, most subjects will have been covered, and for very large $N$ the fact becomes essential that Shakespeare's English is more rich than that of average newspaper writers.

As an alternative strategy, suitable for extremely large data sets, we can truncate the tree at a certain depth $n$. Thus, all branches of the suffix tree are of length $\leq n$ (original tree, Fig. 6f) resp. $\leq n-1$ (modified tree, Fig. 7). Thus, the required memory becomes independent of the text length, and is comparable to that for block entropies of length $n$. Nevertheless, the estimates $\hat{h}_n = \lim_{N\to\infty} \hat{h}_{N,n}$ obtained with such truncated trees are different from the block entropy estimates of Sec. II. In contrast to the latter, they are strict upper bounds to the true block entropies. Using the same data as in Table I, we obtained $\hat{h}_4 = 2.268$, $\hat{h}_5 = 1.982$, and $\hat{h}_6 = 1.851$. This is compatible with Table I. It shows that $h_6$, and thus also $h$, is indeed less than 2 bits/character. Thus the truncated tree yields, given sufficiently large data sets, better exact estimates than the full tree. But it seems not easy to extrapolate these estimates to $n \to \infty$.

To illustrate the working of our method and its main strengths and weaknesses, we finally present some examples showing how the forecasting inherent in the gambling strategy works in detail. We will present two different kinds of forecasting strategies. In both cases, we first trained the algorithm to ''learn'' the structure of some long text. Then we present it with a new short piece of the same text, and ask it to continue it.

In the first test, we always accept the letter with the largest $\hat{p}$, and continue with it, irrespective whether it is right or wrong. In this way we generate random texts which have similar statistics as the original one, but will differ from it due to ''errors'' in estimating the probabilities. In the second test, we just register the letters with largest $\hat{p}$, but we continue with the original text anyhow. In this way we see for each letter how stringent the forecast was, and how correct it was if it was stringent. For all these tests, we used the LOB corpus.

Three artificial texts produced by the first method, after training on texts of increasing lengths, are the following:

Trainings set $10^5$ letters:

ON GATHE PRESIDE POLAFIC GAMENTUNIST CLE VIGOUGHT WOULDS ALSO THE HAS ON BALL LIKE SEPARATION ALTATESTIONS OPPOSED FLAMMELL I MUST OBS PART ONLY CO AND CHANGHAMOST THAT VEYREE TERRY BEATING DAILY THE AS IN IS A IN WERE ROBERT HAND TALKS GVISED HE ISSTROY THESE MORE GENCYERSHI INVE BING PARD LEAD AND ALLORKEOPLE IS NEGOTIATELY AIR LEFT PLACED A AIMEDITATION IN AFTER NEGRESFALLE AND NO GOOD.

Trainings set $10^6$ letters:

ONE THOUGH HEC POLICE THAT EVEN OPENLY WITH BY HOT IS IMMIGRANTS DOWN THE FORMANCES AND SEE ALVANGELICAN MR LIFESSIONALISM WHICH OCH PART OF CRED AND CHANGING THE LAST DIENCY AS WHATEVER VENGE DRAL ARE BUT THE CIVIL FORTUNATELY TOKOU OF THANHAM HICK PRIVATE THE COMMENTARY S HAND BLOWERED WITH THE EASY DALMATTHAT LADIEFENBOW SIR MAIDEN HALL BE SMALLY INCLE A CANNOT A PROFESSION IN HE HAD LABORISM.

Trainings set $5.5 \times 10^6$ letters:

FAILURE LINES AND HAD PAIN IN THE CAPACITIES OF A QUICK OF THE MAJOR ALL ASPECT TO LEAVE THEM AND ITALIAN REWARDS TO TWO SUCH AS MY CERTIFICATES AGAIN MANY ENOUGH <u>DRAUGHINGS</u> BEEN THERE ARE NEITHER <u>BANDA</u> AND MINUTES FOLLOWED WITH THE STAMPS BOX PUT IT GILLINGED CHANGED FOR THE GOING IN ACQUIRED EAR THE EXACT AT JUDGES BETWEEN THE EVENING THE MADE AND THAN HIS BREAK TOUR OF AS THE SOCIAL IN MONARCHING RIVER.

It seems that at least for the latter case the algorithm is nearly perfect on the orthographic level. Using the UNIX spell checker on the last example, it gave only two error messages (see the underlined words). But it seems that it hardly learned any grammatical rules, and any ''meaning'' related to correlations with even longer range was lost completely.

The same conclusion that intra-word constraints are much easier learned than constraints between words can be drawn from the second kind of forecasting test. In Table II we show the letters with the highest estimated probabilities. We see that letters at the beginning of words are much harder to predict than the following ones. This suggests again that the algorithm had the most difficulty in learning interrela-tions between words, while it had learned easily how to continue a word once its first few letters had been given.

TABLE II. After training the tree on the entire LOB corpus, we presented to it the text shown in the first lines. In the lines below it, we show those letters which had estimated probabilities $\hat{p} \geq 0.6$ (bold face), $\geq 0.25$ (sans serif), and $\geq 0.1$ (italic).

| T | H | E | Y | | W | E | R | E | | A | L | L | | A | R | M | E | D | | W | I | T | H | | R | I | F | L | E | S | | A | S | | T | H | E |
|---|---|---|---|---|---|---|---|---|---|---|---|---|---|---|---|---|---|---|---|---|---|---|---|---|---|---|---|---|---|---|---|---|---|---|---|---|---|
| *a* | **O** | **E** | _ | _ | *a* | **E** | **R** | **E** | _ | *a* | **L** | **L** | _ | **T** | **N** | **E** | **Y** | **D** | _ | **T** | **I** | **T** | **H** | _ | **T** | **E** | **G** | **L** | **E** | **S** | _ | *a* | **N** | _ | *a* | **H** | **E** _ |
| *i* | **H** | *a* | | | *h* | *i* | | | | *t* | *s* | *s* | | *o* | | | *s* | | | *a* | *h* | | | | | *a* | *a* | *c* | | | | *m* | *o* | | *r* | | *t* **o** |
| *o* | | | | | *w* | *o* | | | | | | _ | | | | | _ | | | *i* | | | | | | | *o* | *s* | | | | _ | *t* | | | | *i* |
| *t* | | | | | | | | | | | | | | | | | | | | | | | | | | | | | | | | | | | | | |
| N | E | A | R | E | S | T | | O | F | | T | H | E | M | | C | A | M | E | | R | O | U | N | D | | T | H | E | | S | H | A | C | K | | |
| *s* | **E** | **W** | **R** | **E** | **R** | **T** | _ | *i* | **F** | _ | **T** | **H** | **E** | _ | _ | *t* | **O** | **N** | **E** | _ | *t* | **E** | *o* | **N** | **D** | _ | **T** | **H** | **E** | _ | *s* | *a* | **O** | **D** | **K** | | |
| | *a* | *a* | | _ | **S** | | | *o* | *n* | | | | | | | *a* | *a* | *r* | *p* | | *a* | *a* | *u* | *g* | | | *a* | *o* | *a* | | | *t* | **E** | **R** | *h* | | |
| | *o* | *e* | | | | | | *t* | | | | | | | | *i* | *h* | *l* | | | *o* | *o* | | *t* | | | *i* | | | | | *e* | *a* | *p* | | | |
| | *u* | *v* | | | | | | *a* | | | | | | | | *o* | | *m* | | | | | | | | | | | | | | | | *m* | | | |
| | | | | | | | | | | | | | | | | | | *s* | | | | | | | | | | | | | | | | | | | |

This conclusion was confirmed by several subsequent investigations. First of all, we ''scrambled'' the entire text by permuting all words randomly. Thus we generated a surrogate text which had exactly the correct statistics on the word level, but no correct grammatical and syntactical structure at all. We found that its entropy was surprisingly close to the entropy of the original text. Both for the LOB corpus and for Shakespeare, the difference was not more than 0.1–0.2 bits/letter.

Finally, we estimated for every $n$ the average information carried by the $n$-th letter of a word. Here, a word is defined as any string of letters following a blank and ending with the next blank. Results for Shakespeare's collected works, both in the scrambled and in the unscrambled version, are given in Fig. 13 (very similar results were found for the LOB corpus). We see that indeed the first letter in each word carries in average $\approx 6$ times more information than the letters at positions $n \geq 5$! Although some such effect was to be expected, we consider its strength as very surprising. Also, the effect is hardly changed by scrambling.

## VII. SUMMARY AND CONCLUSIONS

We have presented estimates for entropies of symbol sequences drawn from several sources, ranging from chaotic dynamical systems to Shakespeare's works. We believe that our estimates are at least as good as the best ones available in the current literature. In particular, we concentrated on estimates which are based on (implicit) loss-less data compression. For sequences with long range correlations (such as natural languages), these estimates converge very slowly with the sequence length. Hence, realistic estimates of the true entropy requires long sequences and large computational resources. Even if these are available, it seems necessary to extrapolate the estimate obtained for finite length $N$ to $N \to \infty$. We propose an universal ansatz for the latter which seems to work well for all sequences investigated in the present paper.

For written English, our estimates of the entropy converge particularly slowly with $N$. This was to be expected from previous work, and is related to the existence of very long range correlations which are very hard to capture by any algorithm of the sort discussed here. Indeed, ''subjective'' algorithms based on guessing letters by native speakers typically found entropies comparable to ours, and block entropies seemed to converge in these investigations with rates similar to one found in the present paper.

The fact that entropy estimates for finite length texts are too high is related to difficulties in compressing these texts. Indeed, commercial text compression routines (such as UNIX ''compress'') yield considerably worse compression rates. It is also reflected in the very poor ability of the algorithm to produce artificial texts which resemble real texts, as seen in the last section. A more detailed investigation of the source of these problems showed that relationships between words are the main culprit, as they are much harder to learn than orthographic rules. We believe that systematic and detailed investigations such as that shown in Table II and in

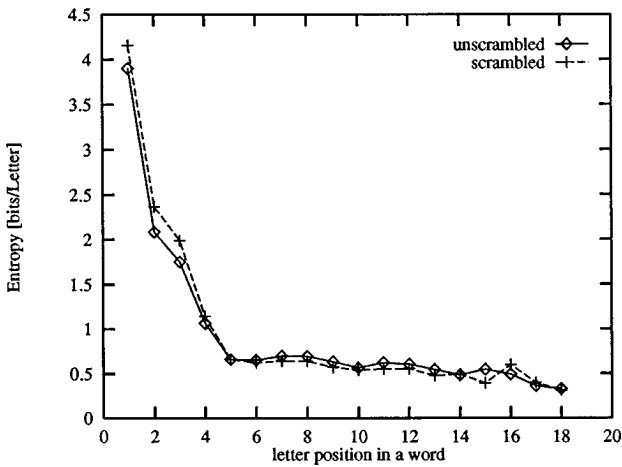

FIG. 13. Average estimated information carried by the $n$-th letter of each word, versus its position $n$. The full line is for the original (unscrambled) version of Shakespeare's collected works, while the dashed line is obtained after scrambling it by permuting its words at random.

Fig. 13 can be extremely useful in developing more efficient entropy estimators and more efficient text compression algorithms.

## ACKNOWLEDGMENTS

This work was supported by DFG within the Graduiertenkolleg ''Feldtheoretische und numerische Methoden in der Elementarteilchen- und Statistischen Physik.''

[1] C. E. Shannon and W. Weaver, *The Mathematical Theory of Communication* (University of Illinois Press, Urbana, IL, 1949).
[2] A. N. Kolmogorov, Dokl. Akad. Nauk. SSSR **119**, 861 (1958); *ibid.* **124**, 754 (1959).
[3] Y. Sinai, Dokl. Akad. Nauk. SSSR **124**, 768 (1959).
[4] P. Grassberger, H. Kantz, and U. Mönig, J. Phys. A **22**, 5217 (1989); G. D'Alessandro, P. Grassberger, S. Isola, and A. Politi, J. Phys. A **23**, 5258 (1990).
[5] P. Grassberger and H. Kantz, Phys. Lett. A **113**, 235 (1985).
[6] C. E. Shannon, Bell Syst. Technol. J. **30**, 50 (1951).
[7] L. B. Levitin and Z. Reingold, preprint (Tel-Aviv University, 1993)
[8] T. M. Cover and R. C. King, IEEE Trans. Inf. Theory **IT-24**, 413 (1978).
[9] N. C. Burton and J. C. R. Licklider, Am. J. Psychol. **68**, 650 (1955).
[10] A. Lempel and J. Ziv, IEEE Trans. Inf. Theory **IT-22**, 75 (1976); J. Ziv and A. Lempel, IEEE Trans. Inf. Theory **IT-23**, 337 (1977).
[11] J. Ziv and A. Lempel, IEEE Trans. Inf. Theory **IT-24**, 530 (1978).
[12] G. G. Langdon, IEEE Trans. Inf. Theory **IT-29**, 284 (1983).
[13] P. Grassberger, IEEE Trans. Inf. Theory **IT-35**, 669 (1989).
[14] P. Shields, Ann. Probab. **20**, 403 (1992).
[15] T. M. Cover, Technical Report **12**, Statistics Department, Stanford University, Palo Alto, CA, 1974.
[16] P. Algoet, Ann. Probab. **20**, 901 (1992).
[17] A. N. Kolmogorov, IEEE Trans. Inf. Theory **IT-14**, 662 (1968).
[18] G. J. Chaitin, J. Assoc. Comput. Mach. **13**, 547 (1966); *ibid.* **16**, 145 (1969).
[19] J. Rissanen, Ann. Stat. **14**, 1080 (1986).
[20] J. Rissanen, IEEE Trans. Inf. Theory **IT-29**, 656 (1983).
[21] T. C. Bell, J. G. Cleary, and I. H. Witten, *Text Compression* (Prentice–Hall, Englewood Cliffs, NJ, 1990).
[22] S. Wolfram, Adv. Appl. Math. **7**, 123 (1986).
[23] P. Billingsley, *Ergodic Theory and Information* (Wiley, New York, 1965).
[24] S. Kullback and R. A. Leibler, Ann. Math. Stat. **22**, 79 (1951).
[25] B. Harris, Colloquia Mathematica Societatis Janos Bolya, 1975, p. 323.
[26] H. Herzel, Syst. Anal. Model Sim. **5**, 435 (1988).
[27] W. E. Caswell and J. A. Yorke *Dimensions and Entropies in Chaotic Systems*, edited by G. Mayer-Kress (Springer-Verlag, Berlin, 1986).
[28] P. Grassberger, Phys. Lett. A **128**, 369 (1988).
[29] D. Ruelle, Ann. NY Acad. Sci. **136**, 229 (1981).
[30] The LOB corpus was provided by Professor D. Wolff, Department of Linguistics, University of Wuppertal, Germany. All other texts were provided as ASCII-texts by Project Gutenberg Etext, Illinois Benedictine College, Lisle.
[31] P. Grassberger, J. Mod. Phys. C **4**, 515 (1993).
[32] E. B. Newman and N. C. Waugh, Inf. Control **3**, 141 (1960).
[33] W. Ebeling and G. Nicolis, Chaos Solitons Fractals **2**, 635 (1992); A. O. Schmitt, H. Herzel and W. Ebeling, Europhys. Lett. **23**, 303 (1993); H. Herzel, A. O. Schmitt and W. Ebeling, ''Finite sample effects in sequence analysis,'' preprint (1993); W. Ebeling and T. Pöschel, Europhys. Lett. **26**, 241 (1994).
[34] H. Kantz and T. Schürmann, Chaos **6**, 167 (1996).
[35] P. S. Laplace, *A Philosophical Essay on Probabilities* (Dover, New York, 1819).
[36] D. H. Wolpert and D. R. Wolf, Phys. Rev. E **52**, 6841 (1995).
[37] M. J. Feigenbaum, J. Stat. Phys. **19**, 25 (1978); **21**, 69 (1979).
[38] J. Rissanen and G. G. Langdon, IEEE Trans. Inf. Theory **IT-27**, 12 (1981).
[39] I. J. Good, *The Estimation of Probabilities: An Essay of Modern Bayesian Methods* (MIT Press, Cambridge, MA, 1965).
[40] R. Krichevsky and V. Trofimov, IEEE Trans. Inf. Theory **IT-27**, 199 (1981).
[41] M. J. Weinberger, J. Rissanen, and M. Feder, IEEE Trans. Inf. Theory **IT-41**, 643 (1995).
[42] F. Willems, Y. Shtarkov, and T. Tjalkens, IEEE Trans. Inf. Theory **IT-41**, 653 (1995).
[43] J. Rissanen, in *From Statistical Physics to Statistical Inference and Back*, edited by P. Grassberger and J.-P. Nadal (Kluwer, Dordrecht, 1994).
[44] M. G. Roberts, Ph. D. dissertation, Electrical Engineering Department, Stanford University, Stanford, CA, 1982.
[45] W. Ebeling and G. Nicolis, Europhys. Lett. **14**, 191 (1991).
[46] Y. Pomeau and P. Manneville, Commun. Math. Phys. **74**, 189 (1980).
[47] S. Hammel, C. K. R. T. Jones, and J. Maloney, J. Opt. Soc. Am. B **2**, 552 (1985).
[48] The binary generating partition for the Ikeda map by use of homoclinic tangencies⁵ was made by H. Kantz.
[49] A. Shenkel, J. Zhang, and Y.-C. Zhang, Fractals **1**, 47 (1993).
[50] M. Amit, Y. Shmerler, E. Eisenberg, M. Abraham, and N. Shnerb, Fractals **2**, 7 (1994).